\documentclass{edm_article}
\usepackage{booktabs}
\usepackage{multirow}
\usepackage{tabularx}
\usepackage{url}
\begin{document}

\title{Automated Feedback in Math Education: A Comparative Analysis of LLMs for Open-Ended Responses}


\numberofauthors{8}
\author{
\alignauthor
Sami Baral\\
       \affaddr{Worcester Polytechnic Institute}\\
       \email{sbaral@wpi.edu}
       \and
Eamon Worden\\
       \affaddr{Worcester Polytechnic Institute}\\
       \email{	elworden@wpi.edu}
       \and
Wen-Chiang Lim\\
       \affaddr{Worcester Polytechnic Institute}\\
       \email{	wlim@wpi.edu}
       \and
\alignauthor
Zhuang	Luo\\
       \affaddr{Worcester Polytechnic Institute}\\
       \email{zluo3@wpi.edu}
\and
Christopher	Santorelli\\
       \affaddr{Worcester Polytechnic Institute}\\
       \email{cjsantorelli@wpi.edu}
\and
Ashish	Gurung\\
    \affaddr{Carnegie Mellon University}\\
    \email{agurung@andrew.cmu.edu}
\and
Neil Heffernan\\
    \affaddr{Worcester Polytechnic Instittue}\\
    \email{nth@wpi.edu}
}

\maketitle

\begin{abstract}

The effectiveness of feedback in enhancing learning outcomes is well documented within Educational Data Mining (EDM). Various prior research has explored methodologies to enhance the effectiveness of feedback. Recent developments in Large Language Models (LLMs) have extended their utility in enhancing automated feedback systems. This study aims to explore the potential of LLMs in facilitating automated feedback in math education. We examine the effectiveness of LLMs in evaluating student responses by comparing 3 different models: Llama, SBERT-Canberra, and GPT4 model. The evaluation requires the model to provide both a quantitative score and qualitative feedback on the student's responses to open-ended math problems. We employ Mistral, a version of Llama catered to math, and fine-tune this model for evaluating student responses by leveraging a dataset of student responses and teacher-written feedback for middle-school math problems. A similar approach was taken for training the SBERT model as well, while the GPT4 model used a zero-shot learning approach. We evaluate the model's performance in scoring accuracy and the quality of feedback by utilizing judgments from 2 teachers. The teachers utilized a shared rubric in assessing the accuracy and relevance of the generated feedback. We conduct both quantitative and qualitative analyses of the model performance. By offering a detailed comparison of these methods, this study aims to further the ongoing development of automated feedback systems and outlines potential future directions for leveraging generative LLMs to create more personalized learning experiences.
\end{abstract}


\section{Introduction}

The growing integration of online learning platforms into traditional educational settings has influenced the development and direction of educational research. The global pandemic, COVID-19, resulted in the adoption of Online Learning Platforms (OLP)\cite{adedoyin2023covid}. Consequently, various OLPs, especially in math education, have gained popularity over the recent years \cite{kim2020exploring}. With the popularity of these platforms, there has been various research investigating effective teaching strategies, with many reporting on the benefit of timely and immediate feedback~\cite{cavalcanti2021automatic, kebodeaux2011defining, dzikovska2014beetle}. Feedback plays a crucial role in facilitating effective learning experiences, offering more than just assessments on the correctness of their answer by providing student-specific guidance. Timely feedback, in particular, can be highly effective in enabling students to rectify misunderstandings, bridge gaps in knowledge, or navigate to subsequent stages of their learning requirements. Prior exploration of effective feedback has reported on the effectiveness of feedback in enhancing learning outcomes, including the use of hints~\cite{phung2023automating}, explanations~\cite{liz2006exemplification}, worked-out examples~\cite{carroll1994using}, and common wrong answer feedback~\cite{gurung2023common, gurung2023identification}, while others caution against the use of certain feedback designs, suggesting that poorly designed feedback can inadvertently impede student progress~\cite{gurung2023identification}.


Automated scoring has been a focus for numerous online learning platforms, with extensive research spanning various fields, including mathematics~\cite{baral2021improving},  writing\cite{mcnamara2015hierarchical, lavoie2020using}, and programming~\cite{price, parihar2017automatic}. The initial works emphasized automating the grading of close-ended questions. However, recent advancements have extended these methodologies to include open-ended problems as well \cite{dikli2006overview}. While early applications of automated scoring primarily focused on augmenting teacher resources in evaluating student responses, more recent explorations have begun to implement these techniques directly within classroom environments \cite{liu2016validation} to support students dynamically in real-time.

The recent advancement and innovation in Large Language Models (LLMs), such as ChatGPT, have introduced a transformative approach to crafting automated feedback systems within educational platforms \cite{Khanmigo}. These developments in LLM technology have demonstrated significant potential in creating diverse mathematical content, providing support for math tutoring, offering detailed explanations, and facilitating the development of automated tutoring systems and educational chatbots that are adept at adapting to a wide range of contextual nuances.

In this study, we delve into the application of pre-trained Large Language Models (LLMs) for both scoring and providing feedback on students' open-ended responses. We particularly assess a fine-tuned LLM derived from Mistral—a Llama variant optimized for mathematics—and compare its efficacy with a leading non-generative model~\cite {botelho2023leveraging}, currently used for the automated assessment of open-ended responses in mathematics. Additionally, we explore how these methods stack up against the capabilities of the GPT-4 model. Given the current limitations on training and fine-tuning GPT-4, we adopt a zero-shot strategy by providing the GPT-4 model with specific rubrics related to the open-ended questions. Toward this, we explore the following research questions:

\begin{enumerate}
\item How does an LLM fine-tuned (GOAT) with a dataset of students' responses and teacher-provided scores compare to the previous state-of-the-art, SBERT-Canberra method in predicting teacher scores for student open-responses?
\item How does the pre-trained GPT4 model compare to the finetuned LLM (GOAT) in the auto-scoring task for open-ended questions?
\item Which of the three models, SBERT-Canberra, GOAT, or GPT4 is preferred for the feedback they generate, according to a detailed assessment protocol and human evaluators with prior teaching experience?
\end{enumerate}


\section{Background} 
\subsection{Open-Ended Problems}
The design of instructional materials for online learning platforms typically falls into two broad categories: close-ended and open-ended problems. Close-ended problems, such as multiple-choice questions, ``check all that apply" scenarios, and arrange in the correct order, are inherently more amenable to automatic grading. The finite range of possible responses in these formats facilitates the development of targeted feedback. On the other hand, open-ended problems include fill-in-the-blanks as well as short and long-answer questions. These types of questions, due to their design, require additional time and resources from the teachers in terms of grading and providing feedback compared to their close-ended counterparts.

While close-ended problems offer the advantage of straightforward automation, they are prone to issues like pattern recognition and guessing~\cite{agustianingsih2019design, papamitsiou2016process}. This susceptibility can sometimes lead to shallow learning and pattern recognition, resulting in the students not engaging deeply with the material. In contrast, open-ended questions are generally considered more rigorous~\cite{klavir2008teaching, hancock1995implementing} and are believed to provide a better gauge of a student's depth of understanding and comprehension of the subject matter~\cite{akay2006, magliano2007, cooney2004}. This depth and rigor, though highly valued in educational contexts, place substantial demands on both the learners, who must articulate their understanding more fully, and the instructors, who face increased burdens in terms of grading and providing meaningful feedback~\cite{Bastin_2003}.

\begin{figure}[!htb]
\centering
\caption{Example of Close and Open-response problem types in mathematics, taken from Illustrative Math Curriculum. Part A,  is a close-ended problem and Part B is an open-ended math problem. } 
\includegraphics[width=\linewidth]
{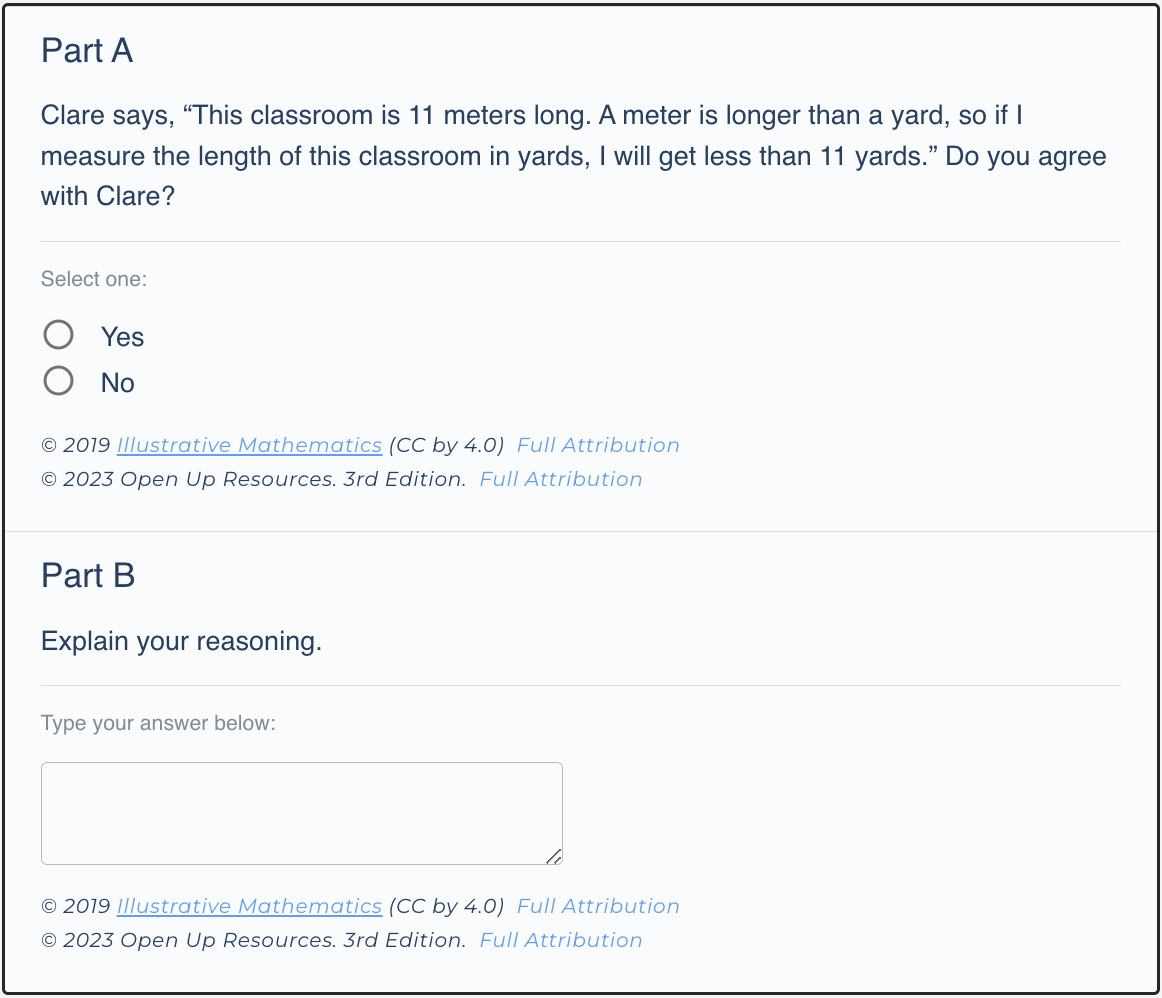} 
\label{fig:open-response}
\end{figure}

OLPs that incorporate mathematics curricula, such as Illustrative Math, EngageNY, or OpenUpResources, available through Open Educational Resources (OER)~\cite{heffernan2014assistments}, typically employ a mix of problem types to enrich the learning experience. This blend aims to leverage the automated feedback and grading capabilities of close-ended problems while the open-ended problems allow teachers to gauge students' understanding and mastery of the subject matter. As presented in Figure~\ref{fig:open-response}, a multi-part problem in Illustrative Maths begins with a close-ended question to confirm basic topic knowledge, followed by an open-ended question designed to evaluate the student's comprehension of the topic. Several researchers and math experts have highlighted the importance of the use of open-ended problems in the effective facilitation of the acquisition and retention of topic knowledge learning process \cite{klavir2008teaching, hancock1995implementing}.


\subsection{Automated Scoring}
There have been many prior works that have focused on automating the assessment (i.e. grading or scoring) of student answers to open-ended questions in the past. Much of this work has leveraged varying approaches that leverage NLP and machine learning. Methods such as C-rater \cite{leacock2003c} use techniques to normalize student responses that vary across syntactic, morphological structure, pronouns, and synonyms to estimate the correctness of student responses to open-ended questions. Other approaches have explored the use of clustering approaches to grade student responses \cite{basu2013powergrading,brooks2014divide}. Other approaches have used deep learning methods that use high-dimensional representations of student work and compare them to exemplar samples, such as in \cite{riordan2017investigating} and \cite{zhao2017memory}.  

Many prior works in NLP have leveraged or expanded upon the idea of creating high-dimensional representations, referred to as embeddings, of student answers (e.g. Word2Vec \cite{mikolov2013efficient} and GloVe \cite{pennington2014glove}). The SBERT-Canberra model\cite{baral2021improving} utilizes a similarity-matching approach using pre-trained SBERT embeddings. Outperforming previous benchmarks in predicting teacher-provided scores for student answers to mathematics open-response problems, this method works by identifying ``similar'' student answers using a measure of Canberra distance \cite{jurman2009canberra} between embeddings; predicted scores are then produced by taking the given score for the most-similar student answer from a pool of historic responses.

\subsection{Automated Feedback}
Automated feedback has been utilized on a variety of domains such as essays, math and coding\cite{borade2021automated, liu2016validation, erickson2020automated}. Hahn et al.\cite{hahn2021systematic} found prior methods have typically attempted automatic at the undergraduate level, but typically reduced bias and assisted teachers by decreasing the time they were required to spend grading. Prior studies in similar domains found that consistent feedback increased student performance in essay writing\cite{kellogg2010does}, and similar studies at the undergraduate level have found feedback decreases math anxiety\cite{nunez2015feedback}, and increases reasoning abilities\cite{fyfe2018feedback}.

A literature review by Deeva et al. \cite{deeva2021review} reviewed automated feedback models used from 2008-2018. They found the majority of automated feedback systems utilized an expert-driven model when providing feedback to students, meaning experts modeled student actions and provided feedback accordingly. We will utilize a data-driven are we utilize existing data to fine-tune our model, whereas many expert-driven models lack substantial data to create models for providing feedback. We will discuss plans to incorporate expert-data in order to make a mixed model in the future, but we intend to first demonstrate that purely data-driven fine-tuned LLMs can outperform SBERT models.\newline


\subsection{LLMs}
Recently, LLMs such as ChatGPT \cite{openaiChatGPT}, Llama2-Chat \cite{touvron2023llama}, and Vicuna \cite{vicuna2023} have emerged as breakthrough technologies in natural language processing, demonstrating impressive performance in language generation and understanding through pre-training on massive text corpora and fine-tuning with specific instructions. Leveraging the capabilities of custom language models for feedback generation, we employ Mistral 7B \cite{jiang2023mistral}, an open-source LLM with outstanding performance on a wide range of benchmarks.

To enhance the applicability of LLMs in downstream tasks, various parameter-efficient fine-tuning (PEFT) techniques have been developed. Notably, \cite{hu2021lora} introduces a low-rank adaptation method named LoRA, which employs rank decomposition matrices in each LLM layer while keeping all pre-trained parameters fixed during fine-tuning. This method, which we also adopt in this paper, aims to optimize the performance of LLMs for specific tasks without altering their original pre-trained parameters.

Additionally, there are also lots of works regarding combining LLMs and education together. \cite{nguyen2023evaluating} evaluates ChatGPT's ability to grade and provide feedback on students' open-ended self-explanations in a digital learning game, finding that while ChatGPT performs well on conceptual questions and generates high-quality feedback, it struggles with decimal place values and number line problems. \cite{hirunyasiri2023comparative} explores the ability of GPT-4 in providing feedback to human tutors on their praise to students in synthetic dialogues, comparing its performance with human graders and identifying areas for improvement. \cite{zhang2023students} investigates the use of ChatGPT in generating formative feedback on Java programming assignments, examining students' perceptions, preferences, and suggestions for improvement. \cite{pankiewicz2023large} explores the use of GPT-3.5 model to generate personalized feedback for programming assignments, demonstrating its positive impact on student performance and learning outcomes, although it may initially lead to over-reliance on the generated hints. However, these works all focus on leveraging OpenAI's models to generate feedback. Our paper attempts to draw inspiration from the concept of CLASS \cite{sonkar2023class}, a design methodology for developing sophisticated Intelligent Tutoring Systems that harness high-performance Large Language Models to offer tutor-like guidance and cultivate interactive conversations. We aim to adapt this approach to fine-tune an open-source Large Language Model for enhanced feedback generation.

\section{Methodology}

In this paper, we present a fine-tuned Large Language Model based on Llama, catered to the dataset of students' open-ended responses in mathematics. We call this model ``GOAT,'' which can generate a numeric score and textual feedback for student open-ended responses in mathematics. We present an in-depth analysis of this model, comparing its performance with the established method, called SBERT-Canberra, from the prior works and the conventional pre-trained GPT4 model. We talk about these methods in detail in the following subsections.\newline

\subsection{Dataset}
For this study, we utilize a dataset from an OLP of students' responses to open-ended math questions with the correctness scores and feedback messages given by teachers to these open-ended responses. We selected a dataset from a specific group of about 50 teachers who used open-ended questions more frequently in their classrooms. 

In preparation for this dataset, we excluded those that featured images within the question, including graphs or tables presented in image format, as our main focus here is on language-based models. Additionally, we eliminated student responses which included images or had images attached to their answers. All HTML tags were stripped from the problems, answers, and teacher feedback messages. Lastly, we converted math symbols, like ``\&ge'', into a more readable format like ``>='' that a language model could interpret.

\begin{table*}[!ht]
    \centering
    \caption{Examples of student open-responses with, teacher-provided feedback and scores to these answers taken from our dataset.}
    \begin{tabularx}{\textwidth}{|X|X|X|X|}
        \hline
         Problem & Student Answer & Teacher Feedback & Teacher Score\\
        \hline
        Explain why 6:4 and 18:8 are not equivalent ratios. & You cannot multiply 4 into 6 and you cannot multiply 8 into 18. & I somewhat see what you are doing but instead you need to see how do you get from 6 to 18 and is that the same scale factor to get 4 to 8. & 1\\
        \hline
        Explain why 6:4 and 18:8 are not equivalent ratios. & They are not equivalent ratios because 6 went into 18, 3 times and 4 went into 8, 2 times & Great job! & 4\\
        \hline
        Write an equation that represents each description. The opposite of negative seven & --7=7 & Great job! & 4\\
        \hline
        Write an equation that represents each description. The opposite of negative seven & 7 & Can you write an equation? & 2\\
        \hline
    \end{tabularx}
    \label{tab:example}
\end{table*}

To train and evaluate our models, we selected 50 random open-response problems, which each had 100 student answers, feedback messages, and teacher scores. We performed an 80-20 train-test split on each question. This meant our training set included 80 student answers, feedback messages, and scores per problem for all 50 problems for a total of 4,000 entries. We then evaluate our models on the remaining 20 students' answers, feedback messages, and scores for a total of 1,000 entries. We compare each model's score to the teacher's assigned score. We also utilize 2 math teachers to manually review 100 random test entries to determine which model performs the best. We had teachers review 2 unique entries for each of the 50 questions in our test set.

We illustrate a few examples of open-ended problems with student responses, teacher-provided feedback, and scores to these responses in Table~\ref{tab:example}. Also, Table~\ref{tab:distribution} presents the distribution of teacher-provided scores within our dataset.


\begin{table}[!ht]
    \centering
    \caption{Score Distribution}
    \begin{tabular}{p{0.25\columnwidth} p{0.35\columnwidth}}
        \toprule
        \centering\arraybackslash \textbf{Score} & \centering\arraybackslash \textbf{Total Responses} \\
        \midrule
        \centering\arraybackslash 0 & \centering\arraybackslash 771 \\
        \centering\arraybackslash 1 & \centering\arraybackslash 768 \\
        \centering\arraybackslash 2 & \centering\arraybackslash 1086 \\
        \centering\arraybackslash 3 & \centering\arraybackslash 816 \\
        \centering\arraybackslash 4 & \centering\arraybackslash 1559 \\
        \bottomrule
    \end{tabular}
    \label{tab:distribution}
\end{table}

\subsubsection{Illustrative Math}
Illustrative Mathematics(IM) is an innovative and widely respected mathematics curriculum designed to deeply engage students in mathematics through problem-solving and interactive learning. IM emphasizes conceptual understanding, problem-solving skills, and critical thinking\cite{IM} and covers a range of grades, typically from Kindergarten through 12th grade. Developed by the combined efforts of both grade K-12 educators and mathematicians, this curriculum utilizes state common core skills to ensure students learn highly valued skills\cite{mccallum2015common} and is considered one of the standard mathematics curricula.

Illustrative Mathematics curriculum widely incorporates open-ended type problems. The online learning platform on which our study is based also adopts this curriculum and follows the same scoring procedure as suggested by this curriculum. As such in our study we utilize the standardized rubric for the assessment provided by Illustrative Mathematics\cite{IM}. This rubric of assessment suggests a 0-4 points rating for student open-responses\cite{IM}. They divide their rubric into 5 tiers which correspond to :
\begin{itemize}
    \item Tier 1 response: Work is complete and correct, with complete explanation or justification.
    \item Tier 2 response: Work shows good conceptual understanding and mastery, with either minor errors or correct work with insufficient explanation or justification.
    \item Tier 3 response: Work shows a developing but incomplete conceptual understanding, with significant errors.
    \item Tier 4 response: Work includes major errors or omissions that demonstrate a lack of conceptual understanding and mastery.
\end{itemize}

We translate this rubric into a 0-4 point scoring system, where a tier 1 would correspond with a 4/4, tier 4 would correspond with a 1/4, and no attempt would correspond with a 0/4 as can be seen in figure~\ref{tab:math-feedback}.

\subsection{SBERT-Canberra}
\begin{figure}[!ht]
\centering
\Description{SBERT-Canberra Model for Automated-scoring and feedback for student open-responses in mathematics, taken from the prior works of Botelho et al.}
\caption{SBERT-Canberra Model for Automated-scoring and feedback for student open-responses in mathematics, taken from the prior works of Botelho et al. \cite{botelho2023leveraging, baral2021improving}} 
\includegraphics[width=.6\linewidth]{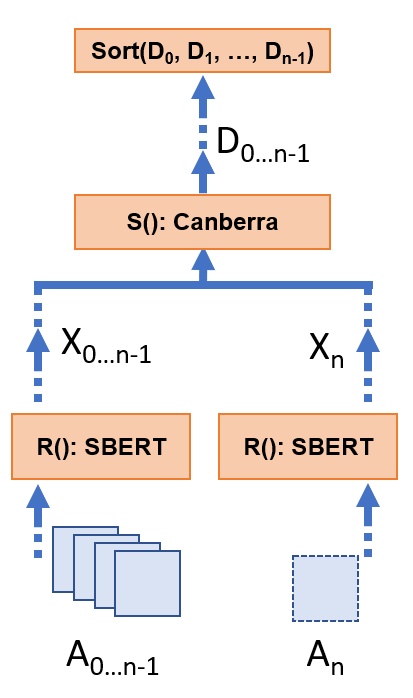} 
\label{fig:sbert-canberra}
\end{figure}

The SBERT-Canbera method from Baral et. al \cite{baral2021improving} presents a similarity-based ranking algorithm for automating assessment for open-ended responses. This method has two parts to it: i) predicting teacher score and ii) predicting teacher feedback for a given student answer. Based on the sentence-level semantic representation of students' open-ended answers, this method presents an unsupervised learning approach as shown in Figure~\ref{fig:sbert-canberra}. The method utilizes a historical dataset collected from an online learning platform, consisting of students' responses with scores and textual feedback from teachers. The model compares any new student response for a math problem,  with the list of responses for the same problem in the historic dataset using sentence-level embeddings from the Sentence-BERT model \cite{nils2019sentence}. Using Canberra distance\cite{jurman2009canberra}, the model finds the most similar answer from the historical dataset to any new student answer and then suggests a score and feedback based on this similar answer. This method is currently in practice in the ASSISTments\cite{heffernan2014assistments} Online learning platform, to recommend scores and feedback suggestions to teachers to give to students' open-responses. 

For our study, we leverage a different dataset than the prior paper, on student open-responses as described in the earlier section. We split the dataset into train and test sets, and use the training data of 50 problems to develop the SBERT-Canberra model and evaluate the results of this model on the test dataset. 

\subsection{GOAT}

\begin{figure*}[!h]
\centering
\caption{The fine-tuning process for the GOAT model for the downstream task of predicting teacher score and feedback for student open-responses in mathematics. } 
\includegraphics[width=.9\textwidth]{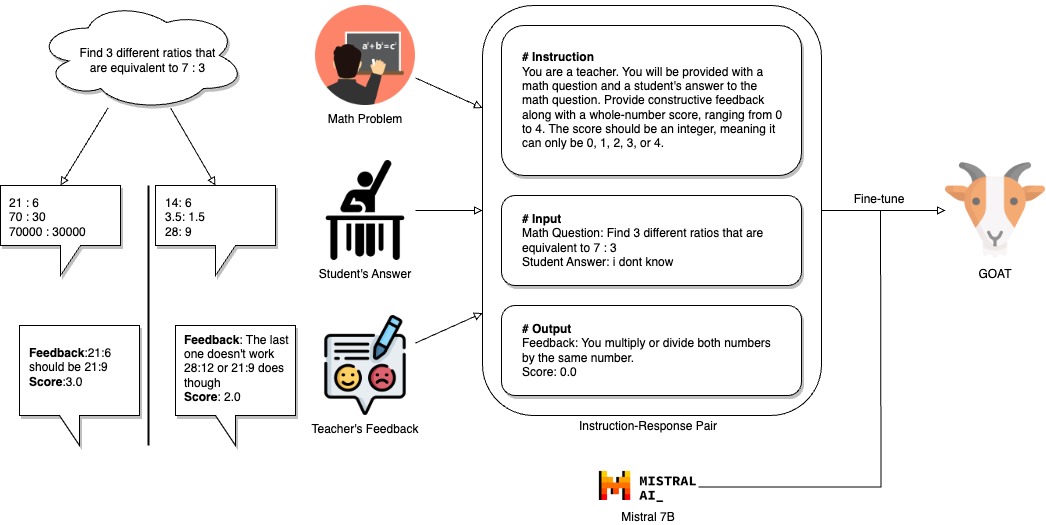} 
\label{Fig.fine-tuning}
\end{figure*}

The GOAT model is our fine-tuned LLM catered to the dataset of student open responses and teacher-provided scores to these responses. 
To develop the GOAT model we fine-tune Mistral 7B\cite{jiang2023mistral}. We fine-tune based on Mistral since it has shown to beat Llama 13B on math, reading comprehension and reasoning. We fine-tuned using LoRA \cite{hu2021lora} since it uses less GPU memory and time and avoids catastrophic forgetting.\newline
To acquire input-output pairs for fine-tuning, we utilize the illustrative grading rubric to design an instructional prompt for each pair, as shown in Figure \ref{Fig.fine-tuning}, amalgamating a math problem and a student's answer into the input, while treating a real teacher's score as the desired output. We utilized 4000 entries data in the training split for fine-tuning and 1000 entries for testing.\newline
Fine-tuning spans 4 epochs with 10 warm-up steps. We initialize the learning rate to 0.0002 and apply a cosine annealing schedule. To address memory constraints, we adopt the gradient accumulation technique, setting gradient accumulation steps to 2, partitioned into micro-batches of 2. The training process, conducted on a single A100 GPU, lasts approximately 2 hours and yields a near-zero loss function when complete.

We determined the optimal inferencing hyperparameters using a validation set of 100 entries which was a subset of the train set. We found argmax$_c$ by finding the parameters which minimized the MSE of our score compared to the teacher score. We found argmax$_c$ to be $temperature$ set to 0.5, $top\_p$ to 0.5, and $top\_k$ to 30.

\subsection{GPT4}

GPT4 is the state-of-the-art language model developed by OpenAI, designed to understand and generate human-like text based on the prompts provided as input to the model. This model has been retained on a diverse and extensive dataset of texts available from the internet, books, and other sources. GPT4 has proven to have significantly improved capabilities in terms of understanding the context, generating relevant text, and handling various complex language tasks.  As such in this work, we explore the applicability of this pre-trained language model in predicting a score and generating appropriate feedback for students' open-response answers in mathematics. 


For our method, we employ the ``GPT4 Turbo'' model, which is the optimized version of the GPT4 model designed to enhance the efficiency and response time and maintain the high quality of the original GPT4 model. With the real-world applicability of this model, being cost and time-efficient, we focus on the use of this version of GPT4 for our study. To explore the performance of the pre-trained model in math assessment tasks,  we employ a zero-shot learning approach with GPT-4, where we do not provide any context examples to the model. For this, we follow a carefully designed prompting strategy, where we provide the model with the problem, the student's answer, and a scoring rubric based on the standard illustrative math rubric. For the prompt engineering process, we followed an iterative approach involving two researchers in math education. We also got high-level feedback from experts with teaching experiences when designing the prompts. They suggested keeping the feedback concise, using middle school-appropriate language, and not giving the answer away through the feedback. 
We incorporated this feedback in our prompts, and the final prompt for the study is shown in Figure~\ref{tab:math-feedback}.

\begin{table}[!ht]
    \centering
    \caption{Final Prompt used as input for GPT-4 model to generate score and feedback for student open-ended responses in mathematics.}
    \begin{tabular}{>{\raggedleft\arraybackslash}p{0.09\columnwidth} p{0.81\columnwidth}}
    \toprule
        \multicolumn{2}{p{0.9\columnwidth}}{You are a middle school math teacher, giving helpful feedback to students on their mathematical reasoning on open-response questions. Keep your feedback direct, under 50 words, and do not give away the answer in your feedback.} \\ \hline
        \multicolumn{2}{p{0.9\columnwidth}}{\textbf{Problem:}} \\
        \multicolumn{2}{p{0.9\columnwidth}}{\{body\}} \\
        \multicolumn{2}{p{0.9\columnwidth}}{\textbf{Student's Answer:}} \\
        \multicolumn{2}{p{0.9\columnwidth}}{\{value\}} \\
        \multicolumn{2}{p{0.9\columnwidth}}{\textbf{Scoring Rubric:}} \\
        1 & Students should get 4 points if their work is complete and correct, with complete explanation or justification. \\
        2 & Students should get 3 points if their work shows good conceptual understanding and mastery, with either minor errors or correct work with insufficient explanation or justification. \\
        3 & Students should get 2 points if their work shows a developing but incomplete conceptual understanding, with significant errors. \\
        4 & Students should get 1 point if their work includes major errors or omissions that demonstrate a lack of conceptual understanding and mastery. \\
        5 & Students should get 0 points if they do not attempt the problem at all. \\
        \bottomrule
    \end{tabular}
    \label{tab:math-feedback}
\end{table}

\subsection{Evaluation}
There are two main parts of the models in this study: i) Predicting a Score and ii) Generating Feedback. Thus, we employ different methods of evaluation for investigating and comparing the performances of these models for the tasks of scoring and generating feedback. 

\subsubsection{Scoring Evaluation}
For the Scoring task, we employ three different evaluation metrics: i) the area under the curve (AUC), ii) the Root mean squared error (RMSE), and iii) multi-class Cohen's Kappa.  Given that the scores for these responses range on a 5-point integer scale ranging from 0 to 4, similar to the prior works\cite{baral2021improving} we employ AUC calculated using the simplified multi-class calculation of ROC AUC, calculating an average AUC over each score category. We use this as the primary metric for evaluating the performance of the models in predicting teacher-provided scores for a given student answer. We employ RMSE which is calculated using the model's estimates as a continuous-valued integer scale, and calculate the multi-class Cohen's Kappa to measure the inter-rater agreement for the scoring task.

\subsubsection{Feedback Evaluation}
For the evaluation of the generated feedback messages, we employ two human evaluators and use a detailed assessment rubric. The evaluators for the feedback messages are the trailing authors of this paper and are Ph.D. students in educational technology with prior teaching experience at the school and college levels. These human evaluators had very little to no information about the models being used in this study. For the study, we randomly sampled a dataset of 100 student answers across 50 different math problems from the test dataset, and the evaluators rated three different feedback messages across the 100 sampled student answers. 

The evaluation rubric is based on: accuracy, relevancy, and the motivational aspects of the feedback messages.

\textbf{Accuracy: } We define accuracy as the factual correctness of the generated feedback. This is a crucial factor for evaluation to ensure that the generated feedback is free of any factual errors. Accuracy is rated on a binary 0-1 scale where a score of 1 indicates the feedback was accurate and a score of 0 indicates the feedback was inaccurate. Further, if feedback was too vague to reflect any factual correctness, the accuracy would be determined by the correctness of the feedback in response to the student's answer. For example, Feedback such as ``Correct Answer!" was deemed accurate only if it accurately reflected the student's performance. 

\textbf{Relevancy: } Relevancy refers to whether or not feedback is relevant to a student's answer or the context of the current problem. For instance, providing a complete problem explanation to every student, regardless of their answer's correctness, would often be deemed irrelevant. Relevancy was also rated on a binary 0-1 scale.

\textbf{Motivation: }  Motivation is rated on a 3-point scale as either -1, 0, or 1. A feedback with motivational aspect would be rated as 1, neutral feedback would be rated as 0 and demotivational feedback would be rated as -1.  For example, responses that said ``Good job", or began with ``Great start, ..." were considered motivating. Responses that were explicitly demotivating, such as ``I am disappointed" were scored as -1, and everything else was scored as 0. 

\textbf{Preferred Model: } In addition to the above-mentioned criteria, the human raters were asked about their preferences on which of the three feedback messages they would pick to give out to their students. 

Finally, at the end of this evaluation, we also asked both teachers to provide their overall recommendations and their perspectives on the quality of feedback generated by the three models involved in this research. 


\section{Results}
\subsection{Scoring Evaluation}
Our comparison of the performance of three advanced models: SBERT-Canberra, GOAT, and GPT-4, in terms of their accuracy in predicting scores provided by teachers, is presented in Table~\ref{tab:model_performance}. These models were evaluated using three different metrics--AUC, RMSE, and Kappa-- to ensure a comprehensive assessment of their predictive capabilities.

Among the three models, the GOAT model outperformed the SBERT-Canberra and GPT-4 models across all three evaluation metrics used. Specifically, the GOAT model achieved an AUC score of 0.7, indicating its strong ability to differentiate between score predictions as ordinal labels. Furthermore, it showed a Root Mean Square Error (RMSE) of 1.119, reflecting its precision in predicting numerical scores, and a Kappa score of 0.422, showcasing 42\% agreement with teacher-provided scores beyond chance.

The SBERT-Canberra model, while not outperforming the GOAT model, had the second-highest AUC score of 0.66 and Kappa of 0.362.
However, it is noteworthy that the SBERT-Canberra model had a higher RMSE of 1.364 compared to the GPT-4 model, which achieved an RMSE of 1.16. Indicating that while the SBERT-Canberra model is relatively strong in predicting the actual scores considering the scores as ordinal labels, it is likely to make more errors on average when considering these scores as continuous values.

The GPT-4 model, with an AUC score of 0.639 and a Kappa score of 0.266, ranked lower in classification performance and agreement with teacher scores compared to the other models. However, its RMSE indicates a relatively moderate level of accuracy in predicting the actual scores, with better performance than that of the SBERT-Canberra model but slightly poorer performance than the GOAT model.

\begin{table}[!h]
    \centering
    \caption{Model Performances on Scoring}
    \begin{tabular}{lcccc}
        \toprule
        \textbf{Model} & \textbf{AUC} &\textbf{RMSE} &\textbf{Kappa} \\
        \midrule
        SBERT & 0.662 & 1.364 & 0.362   \\
        \textbf{GOAT} & \textbf{0.697} & \textbf{1.119} & \textbf{0.422} \\
        GPT-4 & 0.639 & 1.16 & 0.266 \\
        \bottomrule
    \end{tabular}
    \label{tab:model_performance}
\end{table}


\subsection{Feedback Evaluation}
We conducted a qualitative study using two human evaluators to examine the LLM-generated feedback. Both evaluators were experienced teachers with more than 3 years of teaching experience. The evaluation of the feedback generation was conducted on 4 key criteria: accuracy, relevancy, motivation, and overall preference. Accuracy is quantified as the factual correctness of the feedback message, reflecting how often models provided accurate math information. Relevancy represents the applicability of the feedback messages, presenting how often feedback was directly related to the student's answer. Motivation aims to identify whether feedback is motivating, neutral, or demotivating, and preference represents the teacher's preference for the feedback message as judged by our human evaluators.


We defined a message as accurate or relevant if it was unanimously recognized as such by both evaluators. In the aspect of feedback accuracy, the GPT-4 model excelled beyond the SBERT and GOAT models, achieving a consensus accuracy rate of 86\% from both evaluators. In contrast, GOAT and SBERT were deemed accurate 68\% and 52\% times, respectively, by the evaluators. The result for this is presented in Table~\ref{tab:feedback_accuracy}.

\begin{table}[htb]
    \centering
    \caption{Teacher Evaluations for the Accuracy of Generated Feedback}
        \begin{tabular}{lccc}
            \toprule
            Evaluator & SBERT & GOAT & GPT-4 \\
            \midrule
            Teacher 1 & 57 & 69 & 89\\
            Teacher 2 & 85 & 91 & 96\\
            \midrule
            Consensus & 52 & 68 & 86\\
            
            \bottomrule
        \end{tabular}
    \label{tab:feedback_accuracy}
\end{table}

For the relevance of feedback as presented in Table~\ref{tab:feedback_relevancy}, GPT-4 once again outperformed the other two models, securing a unanimous relevance score of 91\% from the evaluators. GOAT was closely behind with 88\% relevance, nearly matching GPT-4, whereas SBERT received a lower relevance score of 76\%.

\begin{table}[htb]
    \centering
    \caption{Teacher Evaluations for the Relevancy of Generated Feedback}
        \begin{tabular}{lccc}
            \toprule
            Evaluator & SBERT & GOAT & GPT-4 \\
            \midrule
            Teacher 1 & 84 & 91 & 93\\
            Teacher 2 & 90 & 96 & 97\\
            \midrule
            Consensus & 76 & 88 & 91\\
            
            \bottomrule
        \end{tabular}
    \label{tab:feedback_relevancy}
\end{table}

In the evaluation of motivational aspects, feedback was categorized as motivating if at least one evaluator found it to be so and similarly categorized as demotivating under the same criterion. The results of the models on the motivational aspect are presented in Tables~\ref{tab:feedback_motivation}  and~\ref{tab:feedback_demotivation}. In terms of motivational impact, the models showed similar performance, with GPT-4 slightly ahead at a 52\% motivation score, followed by GOAT at 48\%, and SBERT at 46\%.

When examining the presence of demotivating elements within the feedback messages, GPT-4 exhibited the least, with only 1\% of its feedback being categorized as demotivating. In comparison, GOAT had a slightly higher incidence of demotivating content at 5\%, and SBERT showed the most, with 11\%, indicating that GPT-4's feedback contained the least demotivational language, whereas GOAT and SBERT utilized more.

\begin{table}[htb]
    \centering
    \caption{Teacher Evaluations for Motivation of Generated Feedback}
        \begin{tabular}{lccc}
            \toprule
            Evaluator & SBERT & GOAT & GPT-4\\
            \midrule
            Teacher 1 & 45 & 47 & 52\\
            Teacher 2 & 40 & 36 & 27\\
            \midrule
            Consensus & 46 & 48 & 52 \\
            
            \bottomrule
        \end{tabular}
    \label{tab:feedback_motivation}
\end{table}

\begin{table}[htb]
    \centering
    \caption{Teacher Evaluations for Demotivation of Generated Feedback}
        \begin{tabular}{lccc}
            \toprule
            Evaluator & SBERT & GOAT & GPT-4\\
            \midrule
            Teacher 1 & 11 & 5 & 0\\
            Teacher 2 & 0 & 4 & 1\\
            \midrule
            Consensus & 11 & 5 & 1\\
            
            \bottomrule
        \end{tabular}
    \label{tab:feedback_demotivation}
\end{table}


In the final analysis of model preference, GPT-4 outperformed the other two models in terms of favorability. On average, the GPT-4 generated feedback was chosen 77.5\% of the time, while both GOAT and SBERT were selected only 12\% of the time each.  
\begin{table}[htb]
    \centering
    \caption{Preferred Model}
    \begin{tabular}{lccc}
        \toprule
        \textbf{Evaluator} & \textbf{SBERT} &\textbf{GOAT} &\textbf{GPT-4} \\
        \midrule
        Teacher1 & 15 & 19 & 69 \\
        Teacher2 & 9 & 5 & 86 \\
        \midrule
        Avg. Percent & 12\% & 12\% & 77.5\%\footnote{There were a few instances in which GOAT and SBERT produced the same feedback. Teacher 1 preferred that feedback in three instances, thus SBERT and GOAT both got a point for being the preferred model in those instances. } \\
        \bottomrule
    \end{tabular}
    \label{tab:preferred_model}
\end{table}

Unsurprisingly, GPT-4 excelled across the entire feedback generation rubric. Its standout performance can be attributed to the advanced capabilities of LLMs like GPT-4 in mimicking human-like fluency and readability, factors that undoubtedly swayed the evaluators' assessments. Consequently, GPT-4's superior proficiency in these areas led it to outshine the others in our qualitative evaluation.


\section{Discussion}
\subsection{Scoring}

In this paper, we introduce the GOAT model, a fine-tuned bespoke solution designed for predicting scores and generating feedback for student responses to open-ended math questions. Our results demonstrate that GOAT outperforms the previous benchmark set by the SBERT-Canberra model in the auto-scoring domain across all three evaluation metrics: AUC, RMSE, and Kappa. However, the model's accuracy, while fair, signals the need for further refinements before full-scale deployment is feasible.

\begin{figure}[htb]
\centering
\caption{Score Distribution of Teachers compared to the three models of SBERT, GOAT and GPT-4 across the test dataset used for the study. } 
\includegraphics[width=0.45\textwidth]{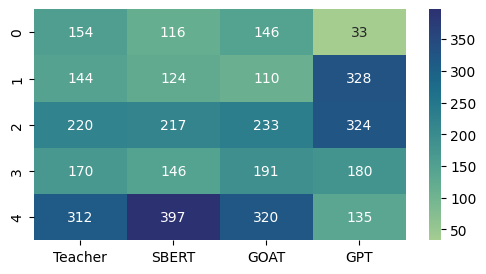} 
\label{Fig.scoring_distribution}
\end{figure}

Notably, both GOAT and SBERT outperformed the GPT-4 model in the scoring task. However, it's crucial to recognize that GPT-4 serves as a generic pre-trained model and hasn't undergone any task-specific fine-tuning or training that both SBERT and GOAT have undergone by utilizing teacher grades on student responses. This distinction is underscored by the alignment between the scoring patterns of SBERT and GOAT with those of the teachers, particularly in their propensity to award scores of 4. In contrast, GPT-4's achieved a distinct score distribution, as illustrated in Figure~\ref{Fig.scoring_distribution}. This variance highlights the nuanced differences in model training and the potential impact on their scoring capabilities. Given that the GPT-4 model utilized a grading rubric from Illustrative Math to assess the quality of student responses, future research should delve into the specific factors contributing to the differences in grading outcomes between GPT-4 and the other models. Identifying the root cause of this grading discrepancy is essential. It could indicate whether the variance is due to teachers' leniency stemming from personalization, a misalignment between the rubric's literal interpretation and teacher expectations in practice, or perhaps a combination of both factors.


\subsection{Feedback}
For the feedback generation task, the GPT-4 model proved to outperform the GOAT and the SBERT-Canberra models across all the evaluation measures as judged by human raters. This highlights the applicability of GPT-4 models in educational platforms for developing automated-feedback systems.  
Further, the proposed GOAT model proved to outperform the prior state-of-the-art SBERT model though both have comparable scores.  

GPT4 proved to be both raters' favorite model, however, GOAT proved to be as good as if not better than BERT. Both teacher 1 and GPT4 ranked GOAT as significantly more accurate and relevant than SBERT. While neither GOAT or SBERT was picked a significant portion of the time, GOAT was rated as accurate 68\% of the time by both raters, whereas SBERT was only rated as accurate by both raters 52\% of the time. GPT was rated as the most accurate being rated as accurate by both raters 86\% of the time. Similarly, GOAT was rated as relevant 88\% of the time compared to 76\% for SBERT, while GPT-4 was still the highest at 91\%.

Teacher 1 considered the following demotivating: ``ask for help!" which occurred 6 times by GOAT, whereas Teacher 2 viewed it as neither motivating nor demotivating. Baral et al. \cite{baral2023investigating} found such phrases can be interpreted as both motivating or demotivating. While Teacher 1 viewed the exclamation mark at the end as demotivating, Teacher 2 felt it was simply encouragement for the student to ask for help from the teacher. One reason for this difference may be the different cultural and teaching backgrounds of the teachers. Teacher 1 was from Singapore, whereas Teacher 2 was from the United States and they showed to have different interpretations of what was motivating and demotivating. Additionally, Teacher 1 taught at the middle school level of mathematics, whereas Teacher 2 taught at the graduate level which may be a cause of their disagreements over which messages would be viewed as motivating, neutral and demotivating.

\subsection{Teacher Opinions}
When asked about their experience, teacher 1 said they viewed the feedback provided by large language models as a helpful starting point, although it sometimes required additional refinement. This teacher likened the experience to receiving an initial draft of feedback from a teaching assistant, which alleviated the cognitive load required to compose feedback from scratch. However, the teacher observed that among the three models, the feedback from GPT was often more thorough, providing deeper insights into the student's work's strengths and areas for improvement. Consequently, this teacher believed that the detailed feedback from GPT was more accessible and easier to tailor for individual needs, in contrast to the feedback provided by the other models.

Teacher 2 said they found the responses provided by each of the three models to be generally accurate, relevant, and motivational. Their assessment indicated a discernible preference for the feedback of GPT. Teacher 2 denoted their preference was influenced by a greater value in the conceptual details and instructions provided by generalized feedback rather than motivational terms which they felt had roots in their individual teaching style. Teacher 2 described GOAT and SBERT as rather stoic in their resultant feedback, which they believed would negatively affect student acquisition of new mathematical concepts and motivation to continue through exercises in practical use. Although the feedback may have been factually correct and concisely conveyed, they felt it would at times be unrelated to the student due to the robotic tone of its responses. They felt GOAT provided a commendable balance between subject matter expertise and motivational nuances which would be necessary given the intended audience. They felt GOAT's responses were often too short, but that a longer more detailed explanation provided in GOAT’s feedback would allow this model to excel.

\section{Limitations and Future Work}

To contextualize the limitations of our study, it is essential to address the primary challenge encountered during the development of the GOAT model, which directly relates to the dataset selection process. The main constraint in training the GOAT model stems from our choice of dataset. We opted to fine-tune the model using the student data that was available to us. However, this led to two main concerns:  


First, a significant challenge was the often subpar quality of teacher feedback. Both SBERT and GOAT models could greatly benefit from detailed and insightful feedback from teachers. However, due to constraints on time and resources, teachers did not always provide feedback that was sufficiently detailed or specific, leading to a collection of training data that lacked depth and variety. This, in turn, diminished the models' effectiveness on test datasets. Specifically, the feedback issues included vagueness, with minimal constructive guidance; irrelevance, where generic comments were applied indiscriminately; and lack of utility, as seen in feedback that merely labeled responses as 'incorrect' without further explanation. Unlike SBERT and GOAT, GPT models, such as GPT-4, are less affected by the quality of these inputs due to their reliance on pre-trained knowledge and the absence of direct training from these specific examples.

Second, the performance of the models varies based on the volume and quality of the data. SBERT, in particular, benefits from a high number of student submissions featuring a wide array of responses and detailed teacher feedback. This environment enables SBERT to closely replicate the discernment of expert teachers, allowing it to detect subtle grading distinctions and expectations for comprehensive answers.

On the other hand, GOAT benefits from high-quality feedback across a variety of problems. As seen in the scoring performance, GOAT has identified patterns in teacher scoring that are not picked up on by SBERT and not captured in the rubric utilized by GPT. Where GOAT performs poorly is its quality of feedback, which mimics the teacher-provided feedback, both good and bad. By training on inaccurate, irrelevant, demotivating, and low-quality feedback messages, the GOAT model captures aspects of teacher feedback that are not as meaningful. While more high-quality training examples on each question would be beneficial, GOAT can benefit from high-quality training examples across multiple questions and use the knowledge it learns for all problems regardless of how many times it has seen each question.

To understand the discrepancies among our raters, we drew parallels with the study by Gurung et al.,~\cite{GurungStudy}, which investigated the variance in teacher grading of identical student responses, both anonymized and nonanonymized. This comparison revealed significant inconsistencies in teacher evaluations even when grading their own students' work. Specifically, the lower-performing teachers demonstrated only a 22\% consistency rate above chance, while the highest performing teachers reached a 73\% consistency rate above chance. These findings underscore the inherent subjectivity and variability in teacher assessments, where teachers frequently diverge from their own previous judgments. Furthermore, Gurung et al. highlighted a tendency for teachers to grade more leniently when aware of the students' identities, pointing to the influence of contextual factors, particularly those considered meaningful by the teachers, on their grading decisions.
We also recognize that many teachers do not utilize the illustrative rubric when grading. They hold a mental model of their rubric for each score on the problems. Additionally, teachers may provide different feedback to their different students depending on their relationship with the student, among other factors. Additionally, some teachers may be harde graders than others, leading to different teachers providing different scores for the same student's answer.

%


We trained GOAT on a large dataset, however, we believe that GOAT would perform significantly better if trained on a more refined dataset. We attribute the presence of inconsistencies in the teacher scores and the presence of vague, relatively shorter, and sometimes irrelevant feedback messages within the training dataset to have affected the performance of the GOAT model.  Further, we believe that many GPT responses were significantly longer and detailed and, thus, were more helpful to students, as is reflected in the teacher's preferred model. As such, we intend to refine our dataset further for the next iteration of our fine-tuned GOAT model. We believe that training only teachers who spend more time writing personalized feedback and supplementing our dataset with feedback from GPT will allow us to provide high-quality while also capturing a human aspect of feedback provided by teachers. We also believe that fine-tuning on problems regardless of the number of examples for each problem will allow us to capture more variety in math problems and answers. Fine-tuning on over 50 problems, even if we don't fine-tune on a large number of instances per problem, will make GOAT generally better at math and generally better at providing feedback.

Lastly, the divergence in human evaluators' perceptions of what constitutes demotivating feedback points to the subjective nature of educational feedback and the need for careful consideration in future model training to ensure that the feedback is universally constructive and encouraging.

\section{Conclusions}
In this work, we present a fine-tuned GOAT model to generate a score and feedback message for a student's open-ended answers. Comparing this method with the traditional method of automated assessment -- the SBERT-Canberra method and the conventional pre-trained GPT-4 model, we find that this method outperforms both the models in the autoscoring task. However, for the feedback generation, the conventional GPT-4 model beats the other two when evaluated by human raters with prior teaching experience.  

\section{Acknowledgments}
We would like to thank past and current including NSF (2118725, 2118904, 1950683, 1917808, 1931523, 1940236, 1917713, 1903304, 1822830, 1759229, 1724889, 1636782, 1535428), IES (R305N210049, R305D210031, R305A170137, R305A170243, R305A180401, R305A120125, R305R220012, R305T240029), GAANN (P200A120238, P200A180088, P200A150306), EIR (U411B190024 S411B210024, S411B220024), ONR (N00014-18-1-2768), NIH (via a SBIR R44GM146483), Schmidt Futures, BMGF, CZI, Arnold, Hewlett and a \$180,000 anonymous donation. None of the opinions expressed here are those of the funders. 

%
\bibliographystyle{abbrv}
\bibliography{EDM_Article_Submission}

\end{document}